\documentclass[twocolumn,aps,prl,showpacs,amsmath,superscriptaddress,longbibliography,notitlepage]{revtex4-1}
\usepackage{amssymb}
\usepackage{mathrsfs}
\usepackage{graphicx}
\usepackage{float}
\usepackage[caption=false]{subfig}
\usepackage[pdftex,colorlinks=red]{hyperref}
\usepackage{verbatim}
\usepackage{txfonts}
\usepackage{epstopdf}
\usepackage[normalem]{ulem}
\usepackage{xcolor}
\usepackage{bm}
\usepackage[utf8]{inputenc}

\def\blue{\textcolor{blue}}

\def\LLH{\textcolor{cyan}}

\begin{document}
\title{Non-Hermitian boundary spectral winding}
\author{Zuxuan Ou}
\affiliation{Guangdong Provincial Key Laboratory of Quantum Metrology and Sensing $\&$ School of Physics and Astronomy, Sun Yat-Sen University (Zhuhai Campus), Zhuhai 519082, China}
\author{Yucheng Wang}
\affiliation{Shenzhen Institute for Quantum Science and Engineering, Southern University of Science and Technology, Shenzhen 518055, China}
\affiliation{International Quantum Academy, Shenzhen 518048, China}
\affiliation{Guangdong Provincial Key Laboratory of Quantum Science and Engineering, Southern University of Science and Technology, Shenzhen 518055, China}
\author{Linhu Li}\email{lilh56@mail.sysu.edu.cn}
\affiliation{Guangdong Provincial Key Laboratory of Quantum Metrology and Sensing $\&$ School of Physics and Astronomy, Sun Yat-Sen University (Zhuhai Campus), Zhuhai 519082, China}
\date{\today}
\begin{abstract}
{
Spectral winding of complex eigenenergies represents a topological aspect unique in non-Hermitian systems, which vanishes in one-dimensional (1D) systems under the open boundary conditions (OBC).
In this work, we discover a boundary spectral winding in two-dimensional non-Hermitian systems under the OBC, originating from the interplay between Hermitian boundary localization and non-Hermitian non-reciprocal pumping.
Such a nontrivial boundary topology is demonstrated in a non-Hermitian breathing Kagome model with a triangle geometry,
whose 1D boundary mimics a 1D non-Hermitian system under the periodic boundary conditions with nontrivial spectral winding.
In a trapezoidal geometry, 
such a boundary spectral winding can even co-exist with corner accumulation of edge states, instead of extended ones along 1D boundary of a triangle geometry.
An OBC type of hybrid skin-topological effect may also emerge in a trapezoidal geometry, provided the boundary spectral winding completely vanishes.
By studying the Green's function, we unveil that the boundary spectral winding can be detected from a topological response of the system to a local driving field,
offering a realistic method to extract the nontrivial boundary topology for experimental studies.
}
\end{abstract}
\maketitle

{\it Introduction.-}
Non-Hermitian systems can support not only topological phases with boundary states protected by conventional band topology~\cite{hasan2010colloquium,qi2011topological},
but also spectral winding topology of complex eigenenergies which has no Hermitian analogue.
Nontrivial spectral winding generally emerges in many non-Hermitian lattices under the periodic boundary conditions (PBC), 
and vanishes when the boundary is opened (i.e. the open boundary condition, OBC), resulting in the non-Hermitian skin effect (NHSE) where bulk states become skin-localized at the system's boundary~\cite{yao2018edge,Yao2018nonH2D,borgnia2020nonH,zhang2020correspondence,okuma2020topological}.
One of the most noteworthy consequences of the interplay between conventional and spectral winding topology is the breakdown of conventional topological bulk-boundary correspondence \cite{Lee2016nonH,xiong2018does},
which has led to recent extensive investigations of its recovery through several different methods \cite{kunst2018biorthogonal,yao2018edge,yokomizo2019non,yang2020non,yang2020non,Lee2019anatomy,herviou2019defining} and many other exciting phenomena induced by NHSE and spectral winding topology \cite{
longhi2019topological,jiang2019interplay,
song2019non,song2019realspace,
Wanjura2020,xue2021simple,wanjura2021correspondence,
budich2020sensor,wang2021generating,longhi2022self,
mu2020emergent,lee2020unraveling,lee2022exceptional,
guo2021exact,
yi2020nonH,
li2021impurity,li2019geometric,li2020critical,Li2021,liu2021exact2,liang2022anomalous,li2022non}. 

\begin{figure}
    \includegraphics[width=1\linewidth]{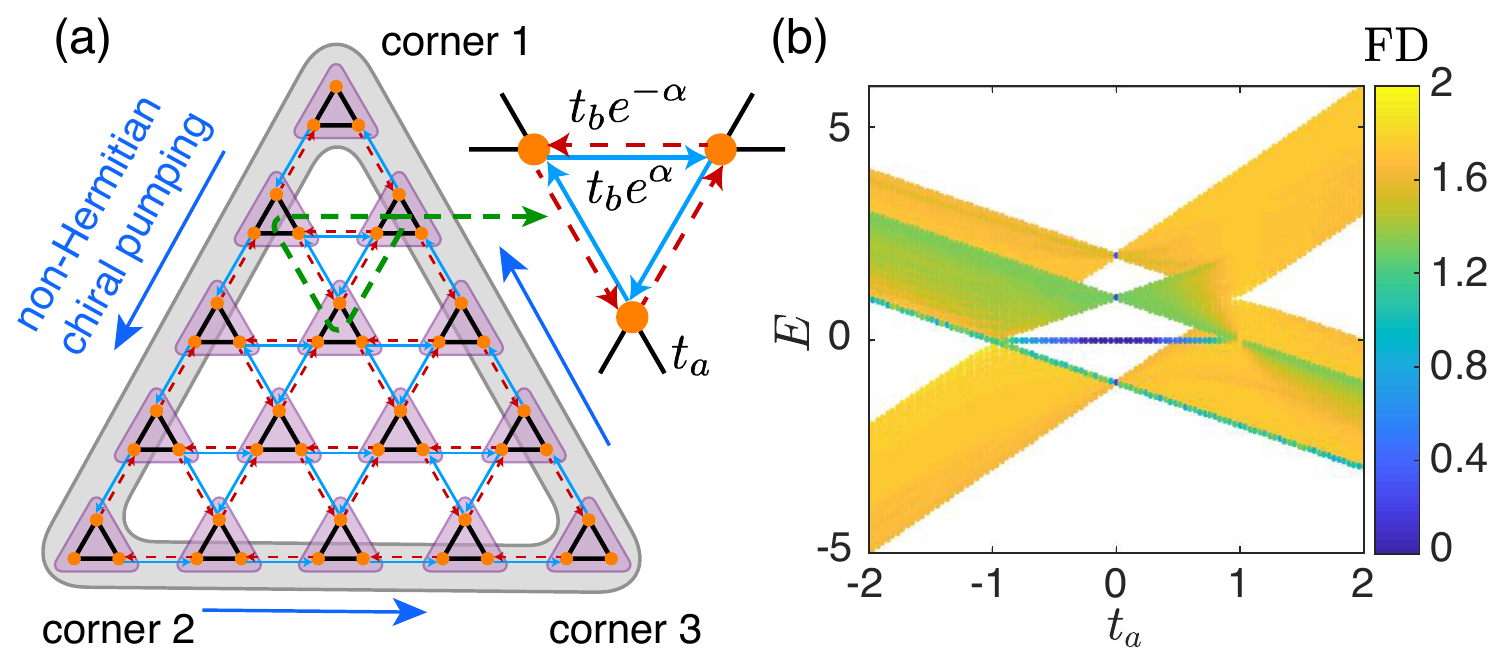}
    \caption{(a) A skecth of the non-Hermitian breathing Kagome lattice with $L=5$ rows of unit cells. Gray area represents the 1st-order boundary of the system, which gives the effective boundary Hamiltonian discussed latter.  Blue arrows indicate the direction of non-Hermitian chiral pumping along the boundary.
    (b) Energy spectrum versus $t_a$, for the system with a triangle geometry in the Hermitian limit of $\alpha=0$. 
    Eigenenergies are marked by different colors according to their corresponding FD.
    Other parameters are $t_b=1$ and $L=30$.
    }
    \label{fig:sketch}
\end{figure}

In contemporary literature, spectral winding topology is mainly studied in one dimensional (1D) systems, as by definition it corresponds to 1D trajectories in the two dimensional (2D) complex-energy plane, which cannot be straightforwardly generalized into higher spatial dimensions.
On the other hand, being a boundary phenomenon,
NHSE in 2D or higher dimension are also far more sophisticated than in 1D, possessing many variations associated with different boundaries and defects due to their richer geometric structures \cite{sun2021geometric,bhargave2021non,schindler2021dislocation,panigrahi2022non,zhang2022universal,jiang2022dimensional}.
In this work, we unveil an exotic aspect of spectral winding in higher dimensions,
namely nontrivial spectral winding for 1D boundary states of 2D lattices under the OBC, 
in sharp contrast to our knowledge of vanishing spectral winding topology of OBC systems.
Its emergence originates from the interplay between topological boundary localization and a non-Hermitian chiral pumping along the boundary, namely asymmetric hoppings with stronger amplitudes toward a chiral direction~[see Fig. \ref{fig:sketch}(a) for an illustration]. 
A similar mechanism is known to be responsible for the hybrid skin-topological effect (HSTE) \cite{Lee2019hybrid,li2020topological,li2022gain,zhu2022hybrid}, a type of higher-order NHSE with topological protection~\cite{kawabata2020higher,fu2021nonH,okugawa2020second}, which induces corner skin-topological localization in 2D lattices.
Interestingly, our example model of non-Hermitian breathing Kagome lattice can support both nontrivial boundary spectral winding and an OBC type of HSTE under the same parameters, but with different triangle and trapezoidal geometries, where the chiral pumping is destroyed by geometric properties of the latter case. 
In the intermedia regime between these two scenarios,
nontrivial boundary spectral winding may even coincide with a weak corner localization, indicating
an anomalous co-existence of the seemingly contradictory boundary spectral winding and HSTE, which require rather different geometric properties along boundaries.
Moreover, we discover that this boundary spectral winding can be detected from a topological response to a driving field when locally perturbing the 2D system, providing a feasible method to extract the nontrivial boundary topology from realistic non-Hermitian systems.

{\it Non-Hermitian breathing Kagome model.-}
We consider a non-Hermitian breathing Kagome model with asymmetric intracell hoppings as shown in Fig.\ref{fig:sketch}(a), with its bulk Hamiltonian given by 
\begin{eqnarray}
    H(\mathbf{k})=\begin{pmatrix}
        0 & t_a +t_b^+e^{-i{k}_1} & t_a +t_b^-e^{i{k}_3}\\
        t_a +t_b^-e^{i{k}_1} & 0 & t_a +t_b^+e^{-i{k}_2}\\
        t_a +t_b^+e^{-i{k}_3} & t_a +t_b^-e^{i{k}_2} & 0\\
    \end{pmatrix},
\end{eqnarray}
where ${k}_1=-k_x/2-\sqrt{3}k_y/2$, ${k}_2=k_x$ and ${k}_3=-k_x/2+\sqrt{3}k_y/2$, $t_b^{\pm}=t_b e^{\pm\alpha}$ represent asymmetric intercell (upward triangle) hopping parameters, and $t_a$ is the amplitude of intracell (downward triangle) Hermitian hopping. Here we set $t_b=1$ as the unit energy. 
In the Hermitian scenario with $\alpha=0$, the Kagome lattice model supports both 1st-order edge states and 2nd-order corner states in certain parameter regimes, as shown in Fig. \ref{fig:sketch}(b) where different bulk and boundary states are characterized by different values of their fractal dimension (FD), defined as
\begin{eqnarray}
{\rm FD}=-\ln[\sum_{\mathbf{r}} |\psi_{n,\mathbf{r}}|^4]/\ln \sqrt{3N},
\end{eqnarray}
with $\psi_{n,\mathbf{r}}$ the wave amplitude at position $\mathbf{r}$ of the $n$-th eigenstate, and $N$ the total number of unit cells.
In a triangle geometry, $N=(1+L)L/2$ with $L$ the number of rows of unit cells in the lattice. 
A 2D bulk state and a 1D edge state in our system shall have their FD close to $2$ and $1$ respectively.
As seen in Fig. \ref{fig:sketch}(b), the breathing Kagome lattice supports 1D edge states (represented by green color for ${\rm FD}\simeq1.2$) in a large parameter regime.




{\it Destructive interference of non-reciprocity and boundary spectral winding.-}
\begin{figure}
    \includegraphics[width=1.0\linewidth]{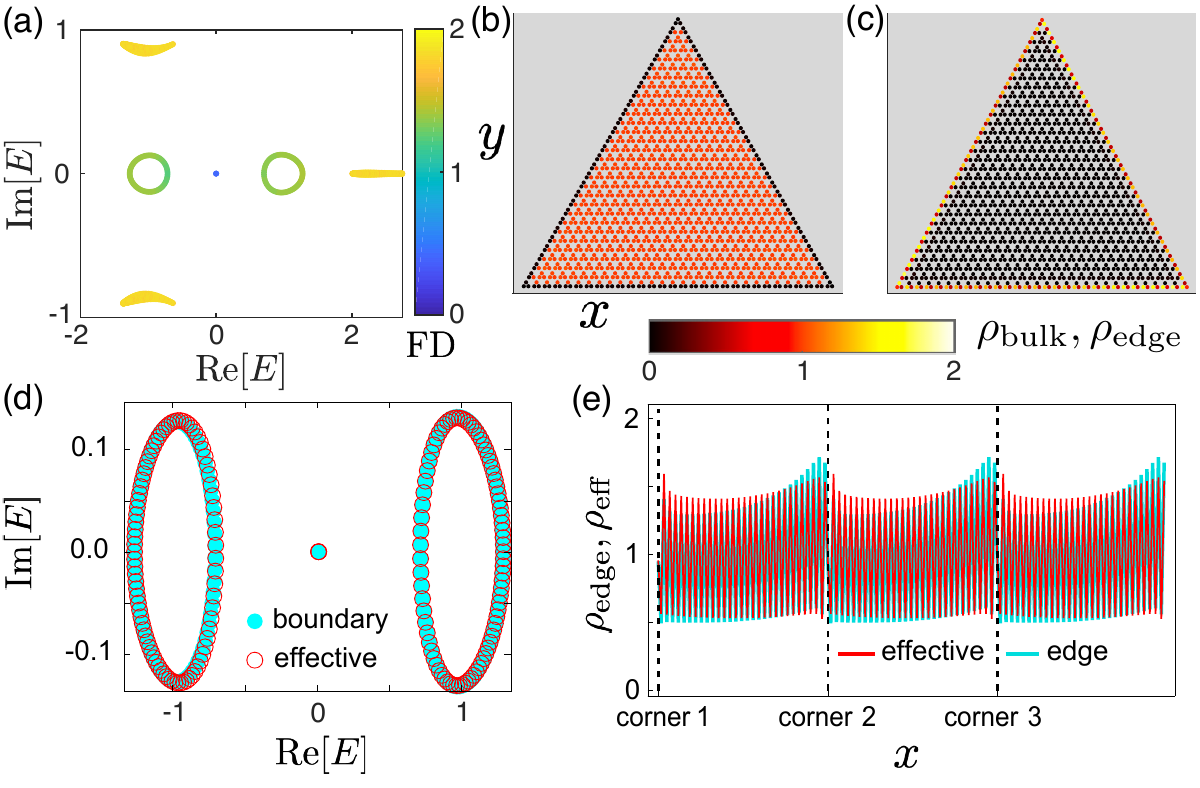}
    \caption{
    (a) Energy spectrum under the OBC, colors indicate the FD of each eigenstate.
    (b) and (c) Summed distribution of bulk states and edge states respectively.
     (d) Spectra of edge states for the 2D lattice (cyan dots) as in (a), and 1D effective boundary system corresponding to the gray area in Fig. \ref{fig:sketch}(a) (red circles). 
     (e) Summed distribution of edge states for the 2D lattice (cyan), and of all eigenstates for the 1D boundary system (red).
     Parameters are $t_a=0.25,t_b=1,\alpha=0.5,L=30$. 
     }
 \label{energy_psi_compare}
\end{figure}
By construction, the model can be viewed as a combination of three sets of non-Hermitian Su-Schrieffer-Heeger (SSH) chains \cite{yao2018edge,SSH,lieu2018topological,Yin2018nonHermitian} along different directions.
Specifically, the three non-Hermitian SSH chains are chosen to be identical, resulting in a $C_3$ rotation symmetry of the system, as shown in Fig. \ref{fig:sketch}(a). 
In this way, the asymmetric hoppings along the three directions form a closed loop and balance out in each unit cell, 
leading to a destructive interference of non-reciprocity in the bulk.
The system is thus net-reciprocal even in the presence of asymmetric hoppings.
As seen in Fig. \ref{energy_psi_compare}(a), FD is close to $2$ for eigenstates in three bulk bands (yellow color) for the system with a triangle geometry,
indicating the absence of NHSE for bulk states.
Consistently, the summed bulk distribution, defined as
$$\rho_{\rm bulk}(\mathbf{r})=\sum_{n\in{\rm bulk}}|\psi_{n,\mathbf{r}}|^2$$
with summation runs over all eigenstates in the bulk bands, also distributes uniformly in the 2D bulk [Fig. \ref{energy_psi_compare}(b)].
On the other hand, 1st-order edge states distribute mostly along 1D edges, 
and are subjected to a net non-Hermitian non-reciprocal pumping.
The same mechanism is known to induce the HSTE in different square and honeycomb lattices.
However, in our model with a triangle geometry, the destructive interference of non-reciprocity limits the choices of the non-reciprocal directions and forbids the HSTE. Namely, in the presence of destructive interference of non-reciprocity along three directions,  
a triangle lattice must have chiral non-reciprocal pumping along its 1D boundary,
hence it is impossible to have two edges with non-reciprocity toward their shared corner. Consequently, we anticipate no hybrid skin-topological corner mode to appear in our system. 

As verified in our numerical calculations,
1st-order edge states are indeed extended along 1D edges, as shown in Fig. \ref{energy_psi_compare}(c) by the summed edge distribution
$$\rho_{\rm edge}(\mathbf{r})=\sum_{n\in{\rm edge}}|\psi_{n,\mathbf{r}}|^2$$
with summation runs over all edge states.
Interestingly, a nontrivial spectral winding is seen to emerge for edge states, even when the system is under OBC [Fig. \ref{energy_psi_compare}(a)]. To understand its emergence, we note that 
these edge states inhabit within the edges of a 2D lattice, which form an effective 1D boundary system without an open boundary, analogous to a 1D non-reciprocal system under PBC. 
By taking the edges of the 2D lattice as a 1D system decoupled from the 2D bulk, we find that its spectrum is almost identical to that of the 1st-order edge states of the original 2D system, as shown in Fig. \ref{energy_psi_compare}(d).
As seen in Fig. \ref{energy_psi_compare}(e), their eigenstates display a slightly different but still extended distribution, as shown by $\rho_{\rm eff}=\sum_{n}|\psi^{\rm 1D}_{n,x}|^2$ with $\psi^{\rm 1D}_{n,x}$ the wave amplitude at position $x$ of the $n$-th eigenstate of the 1D boundary system. 
Thus, the origin of nontrivial boundary spectral winding in our model can be qualitatively understood from this 1D boundary system.
Note that in contrast to an actual 1D PBC system, 1D edges of our 2D model possess an extra lattice site in each corner, acting as impurities to the 1D effective model. As a result, we observe weak eigenstate accumulations toward these corners for both the edge states of the 2D model, and for the effective 1D boundary system [Fig. \ref{energy_psi_compare}(e)]. 

{\it Trapezoidal lattices and an OBC type of hybrid skin-topological localization.-}
\begin{figure}
    \includegraphics[width=1.0\linewidth]{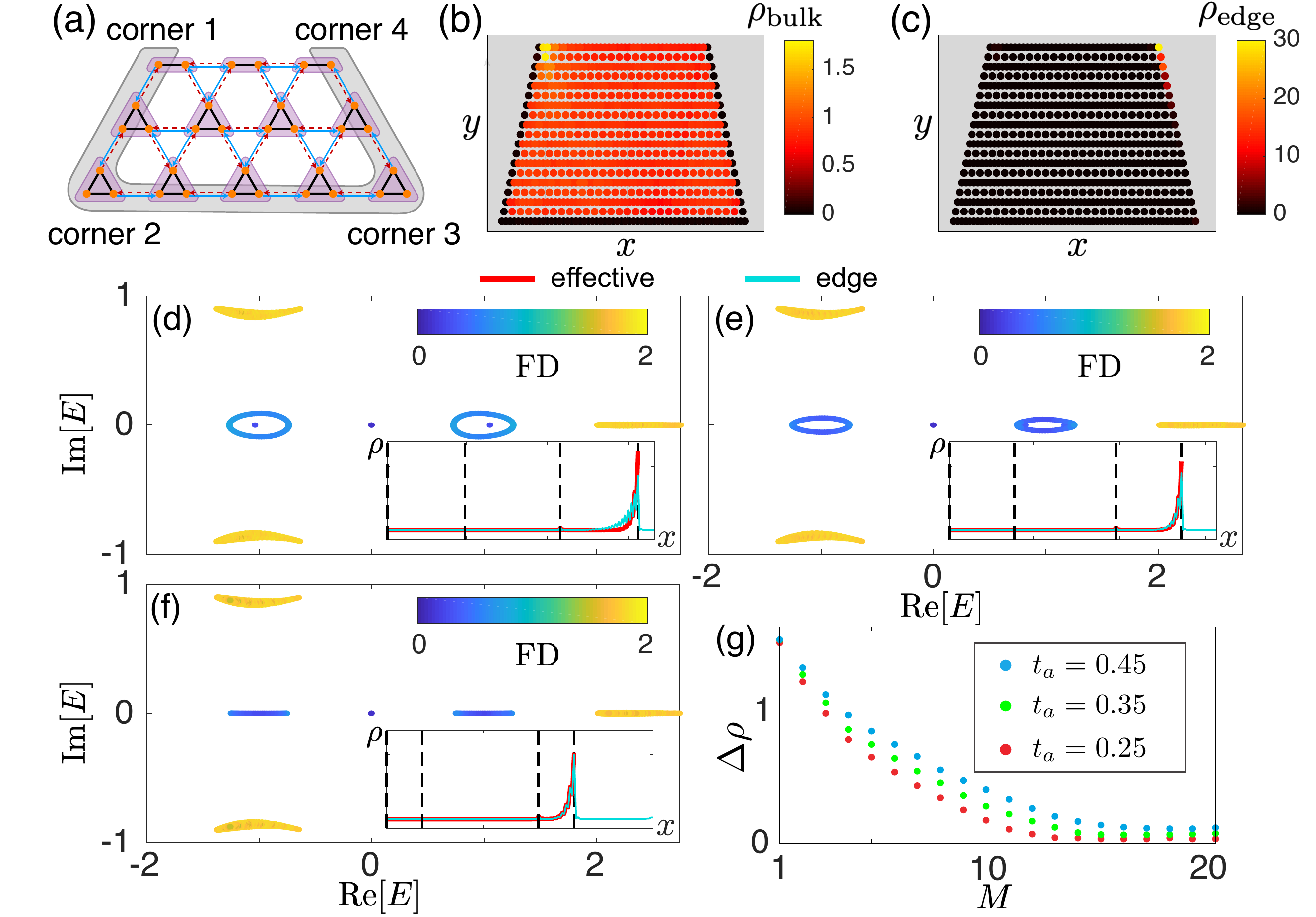}
    \caption{(a) A sketch of a trapezoidal lattice of the non-Hermitian breathing Kagome model. 
    (b) and (c) Summed distribution of bulk states and edge states respectively, with $L=30$ and $M=20$.
    (d) to (f) Energy spectra under OBC with $=30$, and $M=5$, $10$, $20$ respectively, colors indicate the FD of each eigenstate.
    Summed distributions of edge states for the 2D trapezoidal lattice in (d) to (e) are illustrated in their insets  (cyan), together with that of all eigenstates for the 1D boundary system (red). Dash lines correspond to corners $1$ to $4$ from left to right respectively.
     (g) $\Delta\rho=\sum_{x}|\rho_{\rm eff}-\rho_{\rm edge}|/L_{\rm eff}$ as a function of $M$, for several different values of $t_a$.
     $t_a=0.25$ is chosen in (b) to (f). Other parameters are $t_b=1,\alpha=0.5$.
    }
 \label{fig3}
\end{figure}
In the triangle lattice, the emergence of nontrivial boundary spectral winding relies on the chiral non-reciprocal pumping along 1D boundary.
To further confirm this mechanism, 
here we break the chiral pumping channel
by removing top few rows of lattice sites from the triangle lattice. Specifically, we start from a triangle lattice with $L$ unit cells along its bottom row, then remove top $M$ rows of unit cells, and the top lattice sites of unit cells in the $M+1$ row. A sketch of $L=5$ and $M=2$ is shown in Fig. \ref{fig3}(a).
In the resultant trapezoidal lattices,
the top row of lattice sites 
do not form complete upward unit cells, 
and asymmetric hoppings between these sites have stronger amplitudes toward the opposite chiral direction of the rest three edges, blocking the chiral pumping channel circulating the 2D lattice.

Intuitively, 
two domain-walls are formed between top and rest edges at corners 1 and 4, which shall leads to
the emergence of HSTE in the trapezoidal geometry of our model. Yet its behaviors cannot be simply understood from this domain-wall picture.  
To see this, we may redefine a unit cell as a downward triangle [i.e. the green dash line in Fig. \ref{fig:sketch}(a)].
Then, when excluding the boundary lattice sites of left, right, and bottom edges, the rest part of a trapezoidal lattice is formed only by intact unit cells.
Therefore the top geometric edge of a trapezoidal lattice is more closely related to the physical bulk, and shall be considered as a part of it.
In return, the effective 1D boundary no longer forms a closed loop.
Indeed, as observed in Fig.  \ref{fig3}(b) with $L=30$ and $M=20$, bulk states of the system show vanishing distribution only along the left, right, and bottom edges.
On the other hand, edge states in this system exhibit a clear accumulation at top-right corner (corner 4) for the chosen parameters, in analog to the behavior of a 1D non-Hermitian OBC system with the NHSE. 

To further understand the boundary behaviors, we demonstrate the energy spectra for lattices with different sizes in Figs. \ref{fig3}(d)-(f),
with insets illustrating the distributions of their edge states and of a 1D boundary system under the OBC [i.e. the gray area in Figs. \ref{fig3}(a)]. 
In Figs. \ref{fig3}(d), the top $M=5$ rows of unit cells are removed from a triangle lattice with $L=30$.
The resultant system shows a nontrivial boundary spectral winding in its complex spectrum, yet its edge states already become corner-localized (with FD close to $0$) and accumulate at corner 4. However, their accumulating strength is considerably weaker than that of skin states of the effective 1D OBC system.
Such observations are in analogous to the scale-free localization induced by impurities in a 1D non-Hermitian chain, where a boundary impurity connects the 1D chain head to tail, resulting in an impurity boundary condition (IBC) between the OBC and PBC \cite{li2021impurity}. 
In our trapezoidal lattice, 
the top row of lattice sites can be viewed as impurities connecting the two ends (i.e. corners 1 and 4) of the 1D edges,
mimicking the IBC that gives raise to the scale-free localization.

In Figs. \ref{fig3}(e), we demonstrate results with the same number of unit cells in the bottom row ($L=30$), but removing the top $M=10$ rows from a triangle lattice. Its edge states still form a loop-spectrum in the complex plane, yet their distribution becomes closer to the skin states of the corresponding 1D boundary system under the OBC.
Finally, when further removing more rows from the triangle lattice [$M=20$ in Fig. \ref{fig3}(f)], the boundary spectral winding eventually vanishes, and the distribution of edge states become almost identical to the skin states. 
It indicates that the boundary of the 2D trapezoidal lattice becomes an analog of an 1D OBC chain, and supports an OBC type of HSTE, with a vanishing distribution on the other side of the corner, i.e. lattice sites in the top geometric edge of the trapezoidal lattice. This is because these sites act as a part of the bulk and hence behave as a vacuum for the skin-topological states.
To characterize this geometry-dependent behavior, we demonstrate the difference between $\rho_{\rm eff}$ and  $\rho_{\rm edge}$ in Fig. \ref{fig3}(g), defined as
$$\Delta\rho=\sum_{x}|\rho_{\rm eff}-\rho_{\rm edge}|/L_{\rm eff}$$ with $L_{\rm eff}$ the size of the 1D boundary system.
It is seen that $\Delta\rho$ reaches its minimal value at around $M=15$ for $t_a=0.25$, the parameter chosen in other figures.
With larger $t_a$, edge states of breathing Kagome lattice become less localized along its 1D boundary, and their distributions also slightly diverge from that of the 1D boundary system, as shown by the increasing minimal value of $\Delta\rho$ in Fig. \ref{fig3}(g).
In Supplemental Materials~\cite{SuppMat}, we further demonstrate examples of hexagon lattices,
which also exhibit HSTE and boundary spectral winding similar to trapezoid lattices,
and effective OBC and IBC for their 1D edges can be more clearly identified from a Green's function analysis.



{\it Detection of boundary spectral winding through a topological response.-}
In 1D non-Hermitian systems, it has been recently revealed that nontrivial spectral winding has a one-on-one correspondence to a quantized quantity in the response to an external local driving field \cite{Li2021}.
Since edge states of our 2D model effectively give a 1D boundary system, we expect that such a topological response corresponding to spectral winding also emerges here.
Specifically, The topological response quantity is defined as 
$$\nu_{mn}(\beta)=\partial \ln |G_{mn}(\beta)|/\partial \beta,$$
with $\beta$ describing a perturbation to the amplitude of one intra-cell hopping along the system's 1D boundary $t_a\rightarrow t_a e^{-\beta}$, 
 $G(\beta)=1/[E_r-H(\beta)]$ the Green's function regarding a reference energy $E_r$,
and $(m,n)$ labeling the two sites connected by the perturbed hopping.
We note that to obtain a topological response for the boundary spectral winding, the modified hopping can be chosen arbitrarily from the 1D boundary for the triangle lattice, but cannot be chosen from the top row of a trapezoidal lattice, which belongs to the bulk system~\cite{SuppMat}.
In the following discussion, it is chosen to locate at the center of the bottom boundary. 
Physically, $|G_{mn}(\beta)|$ describes the amplitude of a response at site $m$ to a local driving filed at site $n$~\cite{xue2021simple},
thus $\nu_{mn}(\beta)$ describes the changing rate of this response strength with varying $\beta$~\cite{SuppMat}.

\begin{figure}
    \includegraphics[width=1.0\linewidth]{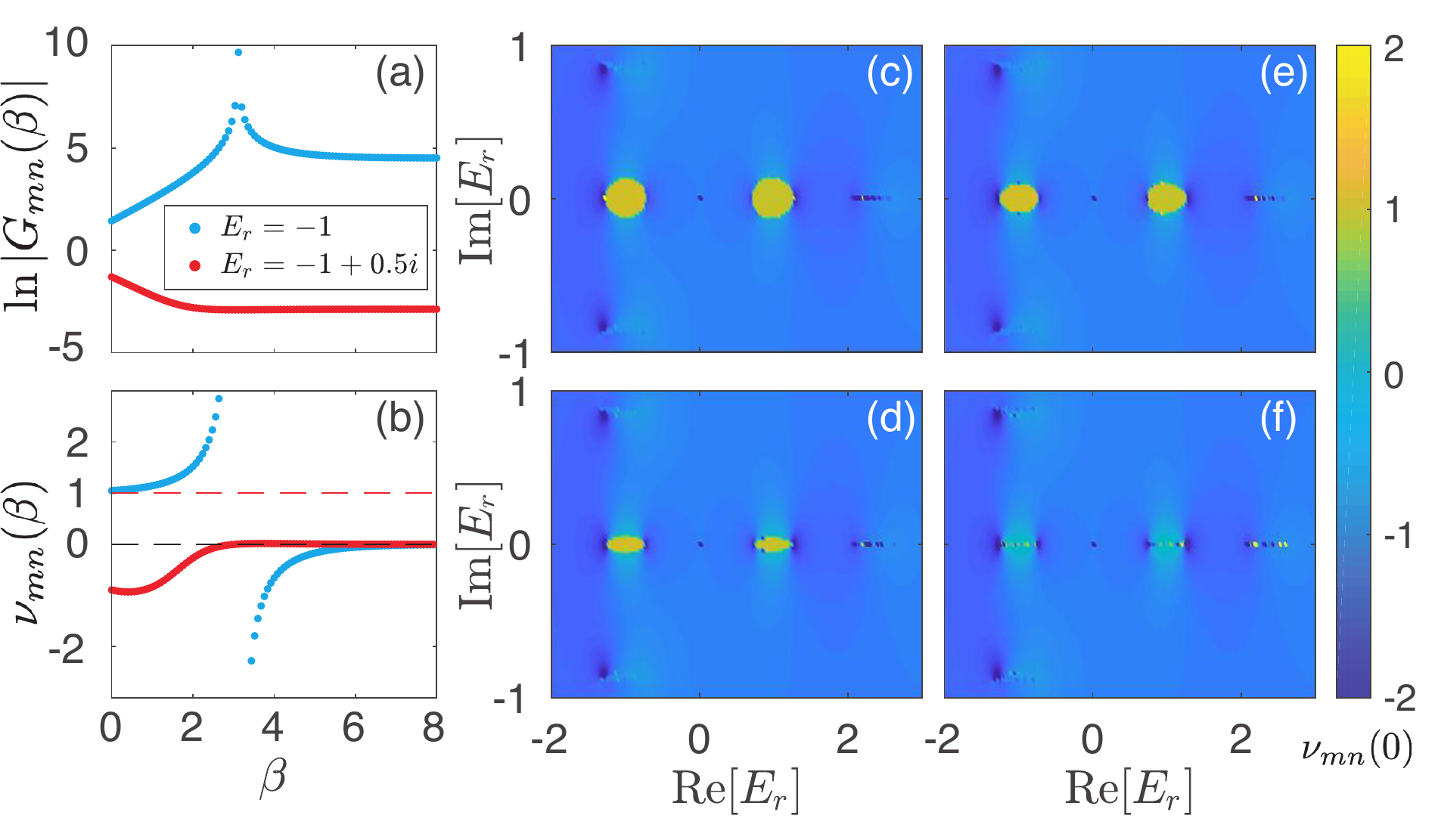}
    \caption{(a) Element $G_{mn}(\beta)$ of the Green's function for the triangle lattice with the same parameters as in Fig. \ref{fig:sketch}(a), for different reference energies $E_r$ enclosed by the loop-like boundary spectrum (blue dots), and within the gap (red dots) respectively.  (b) Topological response $\nu_{mn}(\beta)$ for the same system and parameters as in (a).
    (c) to (f) Topological response $\nu_{mn}(\beta)$ at $\beta=0$ for (c) the triangle lattice with $L=30$ rows of unit cells, and (d)-(f) the trapezoidal lattices with 
for Fig. \ref{fig3}(d) to (f) $M=5,10,20$ rows removed from the triangle lattice respectively.
    In all these results, $m$ and $n$ are chosen as the bottom-left and -right sites of the $15$th unit cell in the bottom row of the lattices.
    }
 \label{fig4}
\end{figure}
We first consider the triangle lattice with the same parameters as in Fig. \ref{energy_psi_compare}.
As seen in Fig. \ref{fig4}(a) and (b), 
$G_{mn}(\beta)$ and $\nu_{mn}(\beta)$ behave rather differently for reference energies inside (blue) and outside (red) the loops of boundary spectrum. 
For the former case, $|G_{mn}(\beta)|$ increases with $\beta$ and eventually stops at a large constant value ($\sim e^{5}$), analogous to the direction signal amplification 1D non-Hermitian system \cite{Wanjura2020,xue2021simple}. In contrast, $|G_{mn}(\beta)|$ decreases to a small value ($\sim e^{-3}$) for a reference energy outside the loops of boundary spectrum.
More importantly, it is seen that $\nu_{mn}(\beta)\simeq 1$ ($\nu_{mn}(\beta)\leqslant0$) for small $\beta$ when $E_r$ is (not) enclosed by the loops of boundary spectrum,
in consistent with the quantized topological response in actual 1D systems~\cite{Li2021,liu2021exact2,liang2022anomalous}.
Therefore we scan $E_r$ for a parameter regime covering the system's full spectrum, and demonstrate the value of $\nu_{mn}(\beta)$ at $\beta=0$ in Fig. \ref{fig4}(c). As seen in the results, the region with nontrivial boundary spectral winding is characterized by $\nu_{mn}(0)\simeq1$, while other regions generally have a non-positive $\nu_{mn}(0)$.
Furthermore, in Figs. \ref{fig4}(e) and (f) we demonstrate the same response quantity $\nu_{mn}(\beta)$ for trapezoidal lattices with the same lattice size and parameters as in Figs. \ref{fig3}(d) to (f).
The nontrivial region is seen to shrink and disappear for trapezoidal lattices when removing more rows of unit cells, which perfectly matches the boundary spectra displayed in Figs. \ref{fig3}(d) to (f).

{\it Conclusion.-}
In summary, we highlight a boundary spectral winding in 2D non-Hermitian systems under the OBC, originating from the interplay between Hermitian (topological) boundary localization and non-Hermitian non-reciprocal pumping. 
The mechanism is similar to that of the HSTE, yet these two phenomena require rather different boundary geometric properties.
Specifically, nontrivial boundary spectral winding naturally arises in a triangle lattice of a non-Hermitian breathing Kagome model, where a destructive interference of non-reciprocity along three directions ensures chiral non-reciprocal pumping along the 1D boundary of the lattice.
On the other hand, a trapezoidal lattice of the same model supports different directions of non-reciprocal pumping for different parts of its boundary, leading to either a boundary spectral winding with a weak corner localization of edge states, or an OBC type of HSTE, depending on the exact shape of the trapezoidal lattice.
In both cases, we find that the boundary spectral winding can be detected from a topological response to a local driving field in the presence of a local boundary perturbation, established from an element of the Green's function matrix associating the response and the driven lattice sites.
Our model is readily for experimental realization with RLC circuit lattices, as it is constructed with the same hopping components as a non-Hermitian 2D lattice already realized in this platform~\cite{zou2021observation}.
The topological response $\nu_{mn}$ can also be detected in such systems by measuring the two-point impedance between sites $m$ and $n$~\cite{Li2021}.
We also note that while this work has focused on a non-Hermitian breathing Kagome model with different geometries, the boundary spectral winding also emerges in other models and geometries, and we have demonstrated several examples in the Supplemental Materials~\cite{SuppMat}.

\bigskip

L. L. would like to thank C. H. Lee, J. Gong, and W. Zhu for helpful discussion. 
This work is supported by the National Natural Science Foundation of China (12104519) and the Guangdong Basic and Applied Basic Research Foundation
(2020A1515110773).

\clearpage

\onecolumngrid
\begin{center}
\textbf{\large Supplementary Materials for ``Non-Hermitian boundary spectral winding''}\end{center}
\setcounter{equation}{0}
\setcounter{figure}{0}
\renewcommand{\theequation}{S\arabic{equation}}
\renewcommand{\thefigure}{S\arabic{figure}}
\renewcommand{\cite}[1]{\citep{#1}}

\section{Hexagonal geometries of the non-Hermitian breathing Kagome model}
In the main text, we have taken a non-Hermitian breathing Kagome model as an example, where nontrivial boundary spectral winding emerges with triangle and some trapezoidal geometries.
More generally, boundary spectral winding can be expected also in other geometries or other 2D non-Hermitian models, as long as chiral non-reciprocal pumping dominates its 1D boundary. In this section, we shall demonstrate extra examples of this model with hexagonal geometries, which can be obtained from
the triangle lattice by removing a few rows of lattice sites from each corner. Specially, we start from a triangle lattice with $L$ unit cells along its bottom row, then remove top $Q$ rows of unit cells, and the top lattice sites of unit cells in the $Q+1$ row on each corner. 


\begin{figure}[ht]
	\includegraphics[width=1\linewidth]{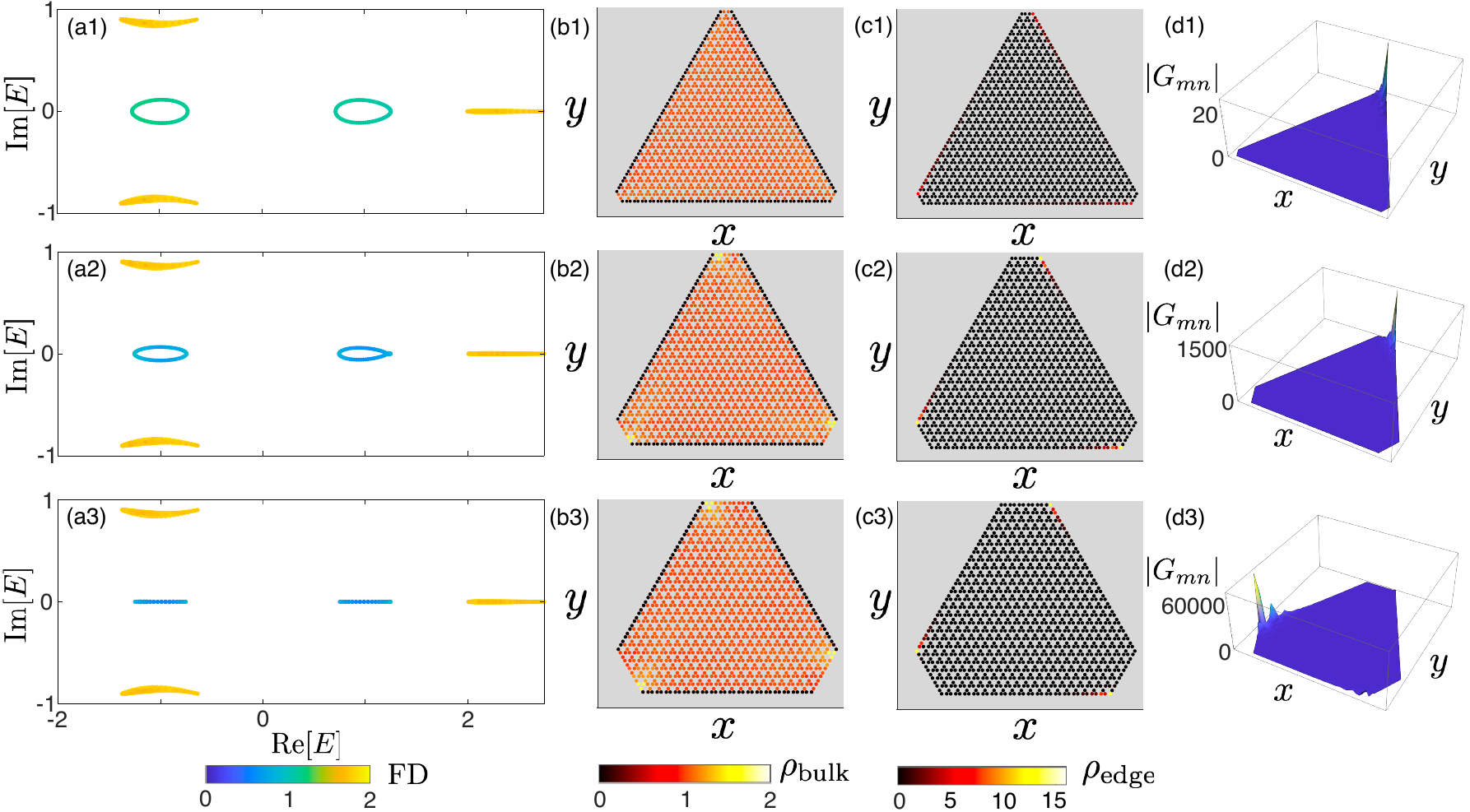}
	\caption{(a1) to (a3) Energy spectra of hexagonal lattices with $L=30$, and $Q=1,3,5$ respectively, colors indicate the FD of each eigenstate. (b1) to (b3) Summed distribution of bulk states in (a1) to (a3) respectively. (c1) to (c3) Summed distribution of edge states in (a1) to (a3) respectively. (d1) to (d3) Element $|G_{mn}|$ of the Green's function for (a1) to (a3). For each case, reference energy is chosen to be $E_r=-1$, which is enclosed by the boundary loop spectrum in (a1) and (a2).
	$n$ is chosen as the top-left corner site for each hexagonal lattice, and $m=(x,y)$ ranges across all lattice sites.
	Other parameters are $t_a=0.25,t_b=1$, and $\alpha=0.5$. 
	}
	\label{sm_hex}
\end{figure}

Removing lattice sites on each corner divides 1D boundary of the triangle geometry into three disconnected segments. 
Similar with the trapezoidal geometry, the top geometric edge of each corner of a hexagonal lattice acts as a part of its physical bulk, 
and bulk states of the system show vanishing distribution only along the other three edges [Figs. \ref{sm_hex}(b1)-(b3)]. 
Edge states in hexagonal lattices are also seen to exhibit a clear accumulation at one side of each of these edges, i.e. three of the six corners of a hexagonal lattice, [Figs. \ref{sm_hex}(c1)-(c3)].
When $Q$ increases, areas with nontrivial boundary spectral winding in the complex energy plane shrink and eventually become some lines [Fig. \ref{sm_hex}(a3)]. 
Meanwhile, 
localization of these corner states also become stronger, which can be seen from larger values of FD [brighter color from Fig. \ref{sm_hex}(c1) to (c3)].
As discussed in the main text, vanishing boundary spectral winding and strong corner localization in Fig. \ref{sm_hex}(a3) and (c3) indicate the emergence of an OBC type of HSTE.


In the main text, we have argued that in a trapezoid lattice, the two ends of its physical edge are (weakly) connected when only a few rows (e.g $M=5$) of lattice sites are removed from the top corner of its parent triangle lattice.
This intuitive picture can be more clearly seen in hexagonal lattices, whose physical edge breaks into three segments.
To see this, we consider the amplitudes of elements $G_{mn}$ of the Green's function $G=1/(E_r-H)$, with $E_r$ a reference energy chosen to be enclosed by the boundary spectrum. As discussed in the main text, This quantity describes the strength of a response field at site $m$ to a driven field at site $n$~\cite{Wanjura2020,xue2021simple}.
Here we have chosen $n$ to be the top-left corner of the hexagonal lattices, and illustrated $|G_{mn}|$ for all sites in the lattices.
It is seen in Fig. \ref{sm_hex}(d1) and (d2) that in the presence of nontrivial boundary winding, $|G_{mn}|$ presents a peak at the top-right corner of the lattices, suggesting that an input signal travels through all edges and is amplified maximally on this corner site.
On the other hand, when nontrivial boundary winding vanishes, $|G_{mn}|$ reaches its maximal value at the bottom-left corner, i.e. the other end of the left physical edge, suggesting that it is disconnected from the rest edges and behaves as an OBC non-Hermitian chain [Fig. \ref{sm_hex}(d3)].
The exponentially increasing  $|G_{mn}|$ also indicates a directional/chiral signal amplification as in an actual 1D system.
The amplification ratio ${\rm Max}[|G_{mn}|]$ is seen to be much smaller in Fig. \ref{sm_hex}(d1), 
which is also in consistence to the understanding that in this case, the edges are more closer to a 1D PBC non-Hermitian chain, which does not support a directional amplification.

\section{Topological boundary response for different geometries}
In the main text, we consider a topological response to detect the boundary spectral winding of our systems, 
namely a response on a boundary lattice site to a local driving field on a neighboring site, with a local perturbation introduced the hopping connecting these two sites.
A one-on-one correspondence is established between the boundary spectral winding and a response quantity defined as
$$\nu_{mn}(\beta)=\partial \ln |G_{mn}(\beta)|/\partial \beta,$$ 
with $\beta$ a parameter controlling the perturbation to a boundary hopping ($t'\rightarrow e^{-\beta}t'$),
$G_{mn}$ an element of the Green's function $G(\beta)=1/[E_r-H(\beta)]$, $E_r$ a reference energy for defining the boundary spectral winding, and $(m,n)$ labeling the response and driven lattice sites respectively.

\begin{figure}[ht]
	\includegraphics[width=1.0\linewidth]{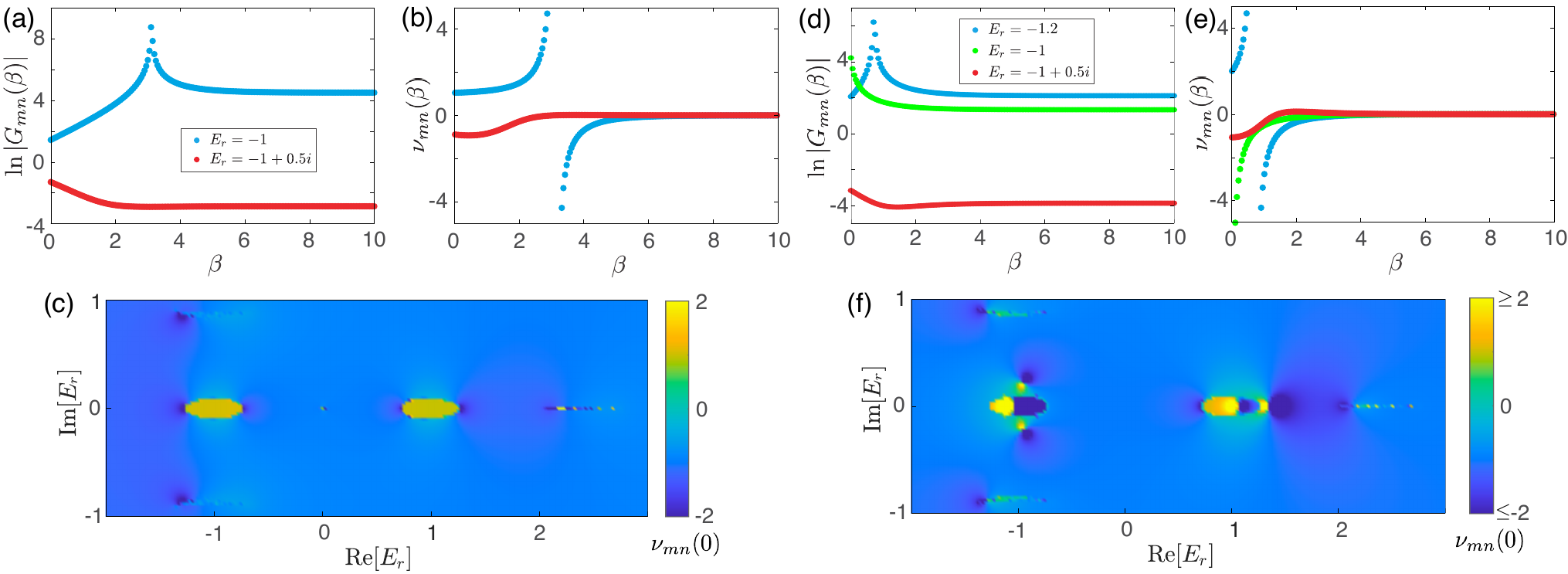}
	\caption{(a) Element $G_{mn}(\beta)$ of the Green's function for the same trapezoidal lattice as in Fig. 3(d) in the main text,
	for different reference energies $E_r$ enclosed by the loop-like boundary spectrum (blue dots), and within the gap (red dots) respectively. 
	The lattice contains $25$ rows of unit cells and $30$ unit cells in the last row, with the top lattice site removed in each unit cell in the top row.
	$m$ and $n$ are chosen as the top and bottom-left sites of the first unit cell in the 13th last row, i.e. two edge sites in the middle unit cell of left edge of the trapezoid lattice. 
	(b) Topological response $\nu_{mn}(\beta)$ for the same system and parameters as in (a). 
	(c) Topological response $\nu_{mn}(\beta)$ at $\beta=0$ for the same system with different $E_r$.
	(d) to (f) the same results as in (a) to (c), only with $m$ and $n$ chosen to be the bottom-right and -left sites of the third unit cell in the top row of the lattices, i.e. two edge sites in the middle unit cell of top edge of the trapezoid lattice. 
	In panel (d) and (e), both green and blue dots correspond to $E_r$ enclosed by the loop-like boundary spectrum.
	Other parameters are $L=30,M=5,t_a=0.25,t_b=1.0,\alpha=0.5$. 
	}
	\label{sm_qr5}
\end{figure}

Intuitively, these two sites can be chosen arbitrarily along edges, except for those in the top edge of a trapezoid lattice, which belongs to the physical bulk of the system.
In Fig. 4 in the main text, the response and driven lattice sites are chosen from the middle unit cell in the bottom of the lattice.
In Fig. \ref{sm_qr5}, we demonstrate the same results as in Fig. 4(c), but with the response and driven lattice sites chosen from left and top edge respectively.
It is seen that the response quantity $\nu_{mn}$ still manifests the boundary spectral winding accurately for the former case, but not for the latter case.

\section{Boundary spectral winding in non-Hermitian Benalcazar-Bernevig-Hughes model}

\begin{figure}
	\includegraphics[width=1\linewidth]{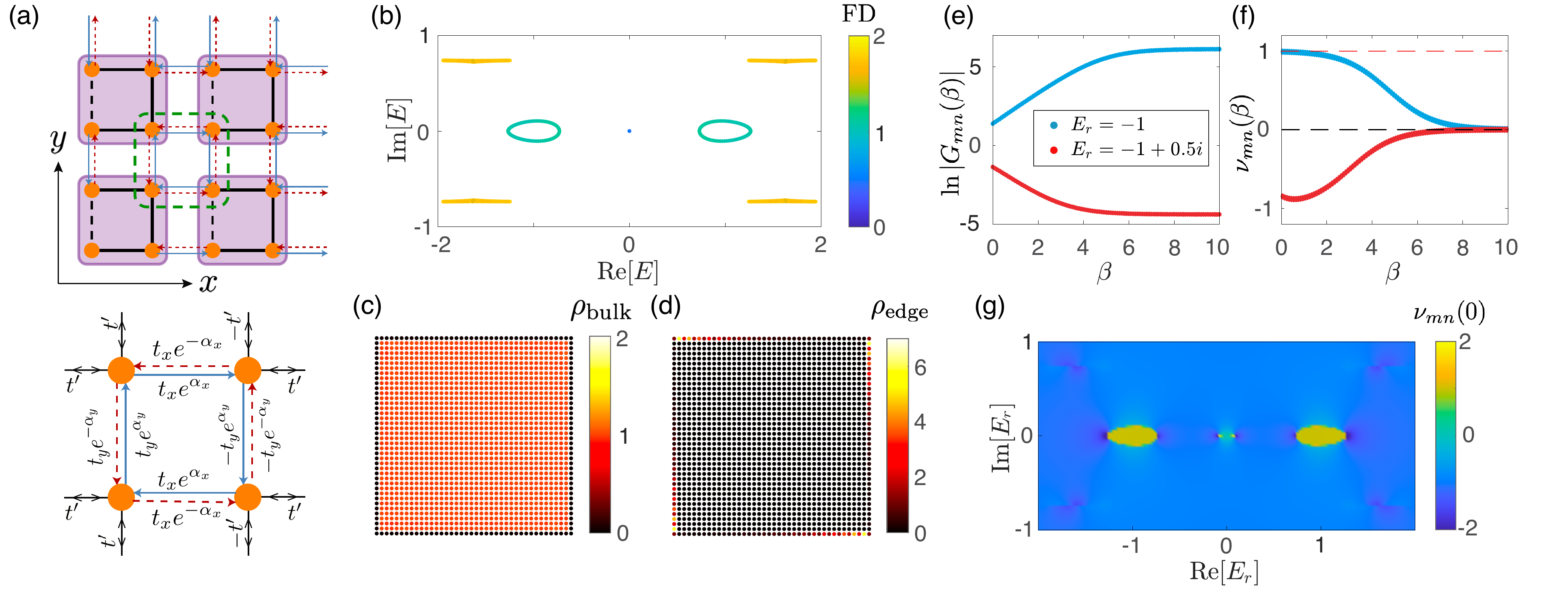}
	\caption{(a) Sketch of the non-Hermitian BBH model. (b) Energy spectra of non-Hermitian BBH model with $L_x=L_y=20$, colors indicate the FD of each eigenstate. (c) and (d) Summed distribution of bulk states and edge states respectively. Other parameters are $t_x=t_y=0.25, t'=1.0, \alpha_x=\alpha_y=0.5$. 
	(e) Element $G_{mn}(\beta)$ of the Green's function of the non-Hermitian BBH model, with $m$ and $n$ chosen as the bottom-left and -right sites of the 11th unit cell in the bottom row of the lattices, and a perturbed hopping parameter $t'\rightarrow e^{-\beta}t'$ between these two sites.
	Reference energies $E_r$ are chosen to be enclosed by the loop-like boundary spectrum (blue dots), and within the gap (red dots) respectively. 
	(f) Topological response $\nu_{mn}(\beta)$ for the same system and parameters as in (f). 
	(g) Topological response $\nu_{mn}(\beta)$ at $\beta=0$. 
	}
	\label{sm_bbh}
\end{figure}

In this section we consider another example, namely the non-Hermitian Benalcazar-Bernevig-Hughes (BBH) model \cite{benalcazar2017HOTI,Benalcazar2017HOTI2} as shown in Fig. \ref{sm_bbh}(a), which also support nontrivial boundary spectral winding in certain parameter regimes.
Its bulk Hamiltonian reads
\begin{equation}
	H(\mathbf{k})=\begin{pmatrix}
		0 & t'+t_x^- e^{-ik_x} & -t'-t_y^+ e^{ik_y} & 0\\
		t'+t_x^+ e^{ik_x} & 0 & 0& t'+t_y^- e^{ik_y}\\
		-t'-t_y^- e^{-ik_y} & 0 & 0 & t'+t_x^+ e^{-ik_x}\\
		0 & t'+t_y^+ e^{-ik_y} & t'+t_x^- e^{ik_x} & 0\\
	\end{pmatrix}, 
\end{equation}
where $t_x^{\pm}=t_x e^{\pm\alpha_x}$ and $t_y^{\pm}=t_y e^{\pm\alpha_y}$ represent asymmetric intercell hopping parameters along $\hat{x}$ and $\hat{y}$ directions respectively, and $t'$ is the amplitude of intracell Hermitian hopping. In the Hermitian scenario with $\alpha_x=\alpha_y=0$, the BBH model supports both 1st-order edge states and 2nd-order corner states. 

The non-Hermitian BBH model we construct can be viewed as a combination of four sets of non-Hermitian SSH chains along two different directions ($\hat{x}, \hat{y}$). In each direction, two different SSH chains are chosen to have opposite non-reciprocity, leading to destructive interference of non-reciprocity in the bulk. FD is close to 2 for eigenstates in four bulk bands (orange color) for the system with a rectangle geometry [Fig. \ref{sm_bbh}(b)], and the summed bulk distribution $\rho_\text{bulk}(\mathbf{r})$ distributes uniformly in the 2D bulk [Fig. \ref{sm_bbh}(c)], indicating the absence of NHSE for bulk states. 
Similar to the non-Hermitian breathing Kagome lattice with triangle geometry in the main text, the non-Hermitian BBH model we construct also has nontrivial boundary spectral winding [Fig. \ref{sm_bbh}(b)],
as its four edges support non-reciprocity toward the same chiral direction. Numerically, we observe
 a weak eigenstate accumulation toward these corners along edges of the 2D system [Fig. \ref{sm_bbh}(d)]. 

To detect the boundary spectral winding, we consider a topological response to an external local driving field, as discussed in the main text.
Specifically, a perturbation is introduced to the amplitude of one intra-cell hopping along the system's 1D boundary, 
\begin{equation*}
	t'\rightarrow t' e^{-\beta}, 
\end{equation*}
and we calculate the Green's function as $G(\beta)=1/[E_r-H(\beta)]$. As shown in Fig. \ref{sm_bbh}(e), 
The topological response associated with boundary spectral winding can be extracted from an element of the Green's funtion $G_{mn}$, with $m$ and $n$ labeling the two lattice sites connected by the perturbed hopping. 
As seen in Fig. \ref{sm_bbh}(e), $G_{mn}$ increases with $\beta$ and eventually stops at a large constant value for a reference energy enclosed by the left loop-like boundary spectrum ($E_r=-1$). In contrast, $G_{mn}$ decreases to a small value for a reference energy outside the boundary spectrum ($E_r=-1+0.5i$). The topological response quantity $\nu_{mn}(\beta)=\partial \ln\vert G_{mn}(\beta)\vert/\partial\beta$ also exhibits distinguished behaviors for different $E_r$, as shown in Fig. \ref{sm_bbh}(f), offering a way to detect the boundary spectral winding. Therefore we scan $E_r$ for a parameter regime covering the system's full spectrum, and demonstrate the value of $\nu_{mn}(\beta)$ at $\beta=0$ in Fig. \ref{sm_bbh}(g). As excepted, the region with nontrivial boundary spectral winding is characterized by $\nu_{mn}(0)\simeq 1$, while other regions generally have a non-positive $\nu_{mn}(0)$.  

%
\end{document}